\documentstyle{article}
\begin{document}
\begin{center}
{\Large \bf Dynamic effects of electromagnetic wave on a damped
two-level atom}
\end{center}

\vspace{.3cm}
\noindent
\begin{center}
{\large Z. Y. Zeng$^{1\ddagger}$, G. J. Zeng$^2$, L. M. Kuang$^2$,
and L. D. Zhang$^1$}
\end{center}

\vspace{.001cm}
\noindent
\begin{flushleft}
{\it $^1$Institute of Solid State Physics, Chinese Academy of
Sciences, P.O. Box 1129, Hefei 230031\\ People's Republic of China \\
\it $^2$Department of
Physics, Hunan Normal University, Changsha, Hunan 410081, People's
Republic of China}
\end{flushleft}

\vspace{.5cm}
\noindent
\begin{center}
{\bf Abstract}
\end{center}

 We studied the dynamic effects of an electromagnetic(EM) wave with
circular polarization on a two-level damped atom. The results
demonstrate interesting  ac Stark split of energy levels of damped
atom. The split levels have different energies and lifetimes, both
of which depend on the interaction and the damping rate of atom.
When the frequency of the EM wave is tuned to satisfy the
resonance condition in the strong coupling limit, the transition
probability exhibits Rabi oscillation. Momentum transfer between
atom and EM wave shows similar properties as the transition
probability under resonance condition. For a damped atom
interacting with EM field, there exists no longer stable state.
More importantly, if the angular frequency of the EM wave is tuned
the same as the atomic transition frequency and its amplitude is
adjusted appropriately according to the damping coefficients, we
can prepare  a particular 'Dressed State' of the coupled system
between atom and EM field and can keep the system coherently  in
this 'Dressed state' for a very long time. This opens another way
to prepare coherent atomic states.

\vspace{1.5 cm}

\newpage

\noindent
\begin{center}
{\bf I. Introduction}
\end{center}

It has been a long history for the study of dynamics of an atom in
Electromagnetic (EM) field.
In the past two decades, there has been
revived and continuous interest to the subject of
atomic motion in a Laser field, due to the variety of
interesting and significant properties displayed in the
interaction between atom and Laser field. These properties have been utilized
to realize the deflection of atom beams by Lasers,
(isotope separation) [$1$]
,  Laser's cooling and trapping of atoms [$2$].
The successful observation [$3$] of the long-predicted Bose-Einstein
condensation in $1995$
was also attributed to the development of the technology
of Laser's  cooling and trapping of atoms.

 The growing theoretical and experimental investigation has been devoted to
 the control of the lifetime of atoms
 in a cavity [$4$], since Purcell proposed the original idea that the
 lifetime of an atom can be modified [$5$]. These studies
 demonstrated  that the spontaneous emission of atom
 can be suppressed or enhanced by a cavity [$6$], and thus the
 lifetime of atom can be controlled artificially.
 On the other hand, efforts to understand the spontaneous emission
 spectrum of two-level atoms when driven by a nearly resonant Laser
 field occupied central stage early on in the development of quantum
 optics.  As is now well known, the primary spectral consequence of
 exposing a two-level atom to a strong driving field is to split
 the undriven atomic emission into a triplet of peaks [$7$]. The atom
 and the driven field can be viewed as a single quantum system termed
 a Photonic Molecular, has eigenstates which has been referred to as
 dressed states [$8$]. As early as 1950s, Lamb et al. investigated
 the shortening of a metastable state caused by the coupling
 between an atom and a static electric field [$9$], and Bethe et al.
 studied such effects for a two-level atom [$10$]. These considerations,
 however, were limited to the static field and excluded the translational
 motion of the atom. Spontaneous emission
 by atom in free space and in cavity was studied recently in the framework
 of density matrix with the help of some approximations as Born-Markov
 approximation [$11$]. It claimed that
 spontaneous emission has no dependence on
 the translational motion of the atom
 in free space, since such dependence would lead to the violation
 of Galilean invariance [$12$]. Cavity mirrors break the Galilean invariance,
 so that atomic eternal motion effects are expected to exist in the
 spontaneous emission of the atom.

 We studied the dynamic effects of a circular-polarized EM wave
 on a  damped two-level atom, taking into consideration
 the center-of-mass motion of the atom. It
 was an  extension of our previous paper concerning the non-damped
 atom in EM field [$13$].  The reason that we use circular-polarized
 EM wave is based on two facts: one is that a plane EM wave can be formed
 from two circularly polarized waves and vice versa, and the other that
 such formulation can voids the use of Rotating Wave Approximation (RWA).
 The procedure adopted is semi-classical, that is to say,
 the atom is treated as
 a quantum system while the electromagnetic wave is considered classically.
 Using such semi-classical procedure, the strict solution of
  Schr$\ddot{o}$dinger equation for the
 damped two-level atom coupling with EM wave can be expected without
 any approximation except the dipole approximation, which is used in
 all the dipole-field interaction. In Sec. {\bf II}, we derived the
 Schr$\ddot{o}$dinger equations for the damped atom and obtained their
  strict solutions.
 Ac stark split of the unperturbed energy levels and the lifetimes of
 the atom in the four split levels were presented in Sec. {\bf III}.
 In Sec. {\bf IV}, we discussed the transition probabilities, finding that
 the coupling between atom and EM wave introduces into the stable
 state such as ground state part of the instability of the other
 states.
 When the frequency of the EM wave was tuned to
 satisfy  resonance condition,
 the transition probability between two unperturbed states
 exhibits Rabi oscillation within a short period
 as in the static electric field case.
In Sec. {\bf V} we calculated the momentum transfer between the atom
and the EM wave under the resonance condition with the consideration
of some limiting cases.
 Sec. {\bf VI} provided a new set of basis
 states-'dressed states',
  which is the superposition of the unperturbed basis states
 of the atom. In this 'dressed state' space, the total
 time-dependent Hamiltonian of the couple system
 can be diagonanized. If an
 EM wave is tuned the
 same frequency as the atomic transition frequency and  its amplitude
 is adjusted appropriately according to the damping coefficients,  then
 we can prepare a dressed state of very long lifetime at which
  the system can keep coherently
  for a quite long time.   Summary along with
concluding remarks was given in Sec. {\bf VI}.

\noindent
\begin{center}
{\bf II. Formulation Of The Problem}
\end{center}

   We first proceed with
   a non-damped two-level atom of mass $m$, transition frequency
   $\omega_a$ and the dipole moment ${\bf D}$ interacting
   with  a circular polarized EM wave propagating in the positive $z$
   direction. The electric field ${\bf E}$ of the $EM$ wave is
   ${\bf E}$=$(E_0\cos(\omega_Lt-kz), -E_0 \sin(\omega_Lt-kz))$, where $E_0$,
   $\omega_L, k$ are the
   amplitude, angular frequency and wave vector respectively. In the
   dipole approximation,
   the Hamiltonian of a two-level atom in an EM wave is $H=P^2/(2m)+
   H_0+H_I$, where $P^2/(2m)$ is the kinetic energy term of the atom,
   $H_0=diag(E_1, E_2)$ is the Hamiltonian associated with internal motion,
   and $H_I=-{\bf D \cdot E}$ is the dipole interaction between
   atom and EM wave.

   we choose the eigenstates  $\mid \varphi_1>$ and $\mid \varphi_2>$
   of the free Hamiltonian $H_0$ whose eigenvalues are, respectively,
   $E_1=\hbar \omega_a/2$, and $E_2=-\hbar \omega_a/2$:
   \begin{eqnarray}
   H_0\mid \varphi_1> &=& E_1\mid \varphi_1>,\nonumber \\
   H_0\mid \varphi_2> &=& E_2\mid \varphi_2>.
   \end {eqnarray}
   Then the dipole transition gives that,
   \begin{eqnarray}
   H^{ii}_I&=&< \varphi_i \mid H_I\mid \varphi_i>=0, \hspace{3cm}
                                i=1,\hspace {0.2cm}  2\nonumber \\
   H^{21}_I&=&<\varphi_2 \mid H_I\mid\varphi_1>=
        -\frac12 \hbar \Omega e^{-i(\omega_L t-kz)}, \nonumber \\
   H^{12}_I&=&<\varphi_1 \mid H_I\mid\varphi_2>=H^{21*}_{I},
   \end{eqnarray}
   where  $\Omega=2DE_0/ \hbar$.
    For a damped two-level atom, it can fall back to lower energy levels
   through spontaneous emission of one or several photons, with
   a probability
   $1/ \tau_i$ per unit time: $\tau_i$ is the lifetime of the unstable
   state $\mid \varphi_i>$. If a state $\mid\varphi_j>$
   is non-damped (such as the ground state)
    in the absence of the coupling $H_I$,
   its lifetime $\tau_j$ is infinite.
   The Hamiltonian for the coupled system between
   a damped two-level atom and an EM wave
    can be readily obtained, by simply adding an imaginary
   term to the corresponding internal energy in the non-damped
   atom's Hamiltonian [$13$]: i. e., $ E_i=E_i-i\hbar\gamma_i /2$,
   where $\gamma_i=1/ \tau_i$ is the damping
   rate of the atom. It is noted that for the Hamiltonian of a
   damped atom in EM field, in which an imaginary term $i\hbar\gamma_i /2$
   exists, then
   it is not hermitian, and thus the probability to find the atom
   in all of its unperturbed states will not always equal to $1$.
   Such system is a dissipative system and its wave functions can not
   be normalized, and the total probability
   of finding the atom in its unperturbed states will not be conserved.
    By solving the Schr{\"o}dinger equation
   $i\hbar\frac{d}{dt}\mid \Psi(t)>=H \mid \Psi(t)>$ in momentum space,
   where $\mid \Psi(t)>=
   \psi_1(t)\mid \varphi_1>+ \psi_2(t)\mid \varphi_2>$,
    we get the following coupled
   equations in momentum space
   \begin {eqnarray}
   i\hbar \frac{d}{dt}\psi_1(p,t)& = &  (\frac{p^2}{2m}+E_1-i\hbar\gamma_1 /2)
                                    \psi_1(p,t)-\frac12 \hbar \Omega
                                    e^{-i\omega_Lt}\psi_2(p-\hbar k,t),
                                    \nonumber\\
   i\hbar \frac{d}{dt}\psi_2(p,t)& = &  (\frac{p^2}{2m}+E_2-i\hbar \gamma_2/2)
                                      \psi_2(p,t)-\frac12\hbar\Omega
                                       e^{i\omega_Lt}
                                       \psi_1(p+\hbar k,t).
   \end {eqnarray}
The coupled Schr{\"o}dinger equations $(3)$ can be decoupled and
solved exactly[$13$]. Assuming that the damped atom is initially
in the state $\mid \varphi_2>$ with momentum $p_0$, then
$\psi_2(p,0)=\delta(p-p_0)$, after some similar algebra as in Ref.
$13$, we obtain the following solutions
\begin{eqnarray}
\psi_1(p+\hbar k,t) & = & \delta(p-p_0)\frac{\Omega}{4\beta(p)}[-
                          e^{-i\omega_1^{+}(p)t}+e^{-i\omega^{-}_{1}(p)t}],
                          \nonumber \\
\psi_2(p,t) & = & \delta(p-p_0)[\frac{\epsilon(p)+
                           \beta(p)}{2\beta(p)}
                          e^{-i\omega_2^{+}(p)t}-\frac{\epsilon(p)-
                          \beta(p)}{2\beta(p)}
                          e^{-i\omega^{-}_{2}(p)t}],
\end {eqnarray}
which can be transformed into coordinate space
\begin{eqnarray}
\phi_1(z,t) & = & \frac{\Omega}{4(2\pi \hbar)^{1/2}\beta(p_0)}
                         \{-e^{\frac{i}{\hbar}
                         [(p_0+\hbar k)z-E_1^{+}(p_0)t]}+
                         e^{\frac{i}{\hbar}
                         [(p_0+\hbar k)z-E_{1}^{-}(p_0)t]} \}, \nonumber \\
\phi_2(z,t) & = & \frac{1}{(2\pi \hbar)^{1/2}}\{\frac{\epsilon(p_0)+
                       \beta(p_0)}{2\beta(p_0)}
                       e^{\frac{i}{\hbar}[(p_0+\hbar k)z+E_2^+(p_0)t]}
                       -\frac{\epsilon(p_0)-\beta(p_0)}{2\beta(p_0)}\times
                        \nonumber \\
            &   &   e^{\frac{i}{\hbar}[(p_0+\hbar k)z+E_{2}^{+}
                         (p_0)t]}\},
\end {eqnarray}
where $ \omega_1^{\pm}(p)=\alpha_1(p)\pm\beta(p),
   \omega^{\pm}_{2} (p)=\alpha_2 (p)\pm \beta(p)$,
   $E_1^{\pm}=\hbar\omega_1^{\pm}, E_2^{\pm}=\hbar\omega_2^{\pm}$,
with
\begin{eqnarray}
   \alpha_1(p) &=& \alpha_2(p)+\omega_L=(E_+/\hbar+\omega_L-i\gamma_+)/2,
   \nonumber \\
   \epsilon(p)   & =& -(E_- / \hbar+\Delta \omega-i\gamma_-)/2,\nonumber \\
   \beta(p)    &=& \sqrt{\epsilon^2(p)+\Omega^2/4},
\end {eqnarray}
with $E_\pm=\frac{(p+\hbar k)^2}{2m}\pm\frac{p^2}{2m}$,
  $\Delta \omega=\omega_a-\omega_L$ and
  $\gamma_\pm=\gamma_1\pm\gamma_2$.

Though the solutions of Schr$\ddot{o}$dinger equation for the damped atom
have the same form as that for non-damped atom, the physical significance
of these solutions are very different.
The crucial discrepancy lies in that the coefficients and the exponential
terms concerning the energy and lifetime in the solutions of
Schr$\ddot{o}$dinger equation for damped atom are
complex function, while they are real function in that
for non-damped atom.  In addition, the wave functions of the atom
can not be normalized, and the probabilities of finding the atom in
its unperturbed states will be not always equal to $1$.

\noindent
\begin{center}
{\bf III. Ac Starc Split Of Energy Levels}
\end{center}

  In the above section, we mentioned that the coefficients and
the exponential terms related to the energy and the lifetime
in the solutions of
Schr$\ddot{o}$dinger equation given by Eqn. ($5$) are both complex
function. They can be decomposed
into two parts-the real part designated by superscript $R$ concerning energy,
and the imaginary part by superscript $I$ is related to
damping rate. After some algebra,
we have
\begin{eqnarray}
E_1^{\pm R} &=& [E_++\hbar\omega_L \pm 2\hbar\beta^R(p)]/2, \nonumber \\
E_2^{\pm R} &=& [E_+-\hbar\omega_L \pm 2\hbar\beta^R(p)]/2, \nonumber \\
E_1^{\pm I} &=&\hbar [\gamma_+ \pm 2\beta^I(p)]/2, \nonumber \\
E_2^{\pm I} &=& \hbar[\gamma_+ \pm 2\beta^I(p)]/2,
\end{eqnarray}
where
\begin{eqnarray}
\beta^R(p)&=&\frac{\sqrt{2}}{4}
\{\sqrt{[(E_-/ \hbar+\Delta \omega)^2+\Omega^2-
\gamma_-^2]^2+4\gamma_-^2(E_-/ \hbar+\Delta\omega)^2}+\nonumber \\
         & &(E_-/ \hbar+\Delta \omega)^2+\Omega^2-\gamma_-^2\}^{1/2},
                \nonumber   \\
\beta^I(p)&=&\frac{\sqrt{2}}{4}
\{\sqrt{[(E_-/ \hbar+\Delta \omega)^2+\Omega^2-
\gamma_-^2]^2+4\gamma_-^2(E_-/ \hbar+\Delta\omega)^2}-\nonumber \\
         & &(E_-/ \hbar+\Delta \omega)^2-\Omega^2+\gamma_-^2\}^{1/2}.
\end{eqnarray}
 Inspecting  Expressions ($4$) and ($5$), we find  that
 both $\phi_1(z,t)$ and  $\phi_2(z,t)$
 comprise two plane waves, in which
 the energies and the lifetimes of the atom are different with the momenta
 being the same. Thus the two unperturbed
 levels of the atom are split into four
 levels of different energies,
  which are given by Eqn. ($7$).
 The energy difference $\Delta E= E_1^+-E_1^-=  E_2^+-E_2^-=2
 \hbar\beta^R(p_0)$
 is the same for the two unperturbed
 levels of the atom. While the lifetimes(damping rate)
 of the
 atom in the four split-levels are obviously modified as compared  with
 that in the unperturbed two states, though the imaginary
 part of the energy term can not be directly regarded as  the inverse of
 the lifetime. In Fig. $1$, we depicted
 the above energy level split caused by a time-varying
EM field is referred to as ac Stark split.

\noindent
\begin{center}
{\bf IV. Transition Probabilities and Rabi Oscillation}
\end{center}

It is well known that, when a two-level atom is interacting with a
static electric field, Rabi oscillation occurs, i. e.,
its transition probability will oscillates
between the two-unperturbed states $\mid \varphi_1>$ and $\mid \varphi_2>$.
From Eqn. ($4$) or ($5$), we can readily obtain the probabilities
 of finding the
damped atom at time $t$ in its unperturbed
state $ \mid \varphi_1>$ and $ \mid\varphi_ 2>$
\begin{eqnarray}
\rho_1(p_0+\hbar k,t) &=& \frac{\Omega}{8\mid \beta(p_0)\mid^2}\{cosh[2
                           \beta^I(p_0)t]-cos[2\beta^R(p_0)t]\}e^{-\gamma_+t},
                           \nonumber \\
\rho_2(p_0,t) &=& \frac{1}{2\mid \beta(p_0)\mid^2}\{[\mid \epsilon(p_0)\mid^2
                   +\mid \beta(p_0)\mid^2]cosh[2\beta^I(p_0)t]
                   +[\gamma_-\beta^I(p_0)-(E_-/\hbar+\Delta \omega)\nonumber \\
              &  &   \beta^R(p_0)]sinh[2\beta^I(p_0)t]+
                    [\mid \beta(p_0)\mid^2-\mid \epsilon(p_0)\mid^2]
                    cos[2\beta^R(p_0)t]-\frac12[\gamma_-\beta^R(p_0)\nonumber \\
              &  &  +(E_-/ \hbar +\Delta \omega)\beta^I(p_0)]
                     sin[2\beta^R(p_0)t]\}e^{-\gamma_+t},
\end {eqnarray}
Eqn. ($9$) demonstrated a fact that the
probabilities of finding the atom in its unperturbed state $\mid\varphi_1>$
or $\mid \varphi_2>$ evolve with time in a quite complicated way.
They depend not only on the amplitude and
frequency of the coupling circular-polarized EM wave, also
on the physical qualities of the atom such as dipole moment and
kinetic energy.
More importantly, the coupling of a two-level atom
with a EM wave introduces into the stable state part of the
instability of the
other state. This can be verified as
we set $\mid \varphi_2>$ to be a non-damped state ( $\gamma_2=0$),
damped exponentials still exists in the probability to find
the atom in this non-damped state $\mid \varphi_2>$.
Then we conclude that when a damped atom is exposing to an EM field,
there exists no longer stable state for this atom.
If the frequency of the EM is tuned such that
$E_-/ \hbar+\Delta \omega=0$ (which we defined as 'Resonance'), then
the above
probabilities have the following simple form
\begin {eqnarray}
\rho_1(p_0+ \hbar k, t) & = & \frac{\Omega^2}{\Omega^2-\gamma_-^2}
                              \sin^2[(\Omega^2-\gamma_-^2)^{\frac12}t/2]
                              e^{-\gamma_+t},\\
\rho_2(p_0, t) &=&  \frac{\Omega^2}{\Omega^2-\gamma_-^2}\{
                              \cos^2[(\Omega^2-\gamma_-^2)^{\frac12}t/2]+
                            \frac{ \gamma_-(\Omega^2-\gamma^2_-)^{\frac12}}
                            {2\Omega^2}
                             \sin[(\Omega^2-\gamma^2_-)^{\frac12}t]
                             \nonumber  \\
               &  &           -\frac{\gamma_-^2}{\Omega^2}\}e^{-\gamma_+t}.
\end {eqnarray}
for strong coupling($\Omega>\gamma_-$), while for
for week coupling, i. e.,  $\Omega<\gamma_-$, we have
\begin {eqnarray}
\rho_1(p_0+ \hbar k, t) & = & \frac{\Omega^2}{\gamma_-^2-\Omega^2}
                              \sinh^2[(\gamma_-^2-\Omega^2)^{\frac12}t/2]
                              e^{-\gamma_+t},\\
\rho_2(p_0, t) &=&  \frac{\Omega^2}{\Omega^2-\gamma_-^2}\{
                              \cosh^2[(\gamma_-^2-\Omega^2)^{\frac12}t/2]+
                            \frac{ \gamma_-(\gamma^2_--\Omega^2)^{\frac12}}
                            {2\Omega^2}
                             \sinh[(\gamma^2_--\Omega^2)^{\frac12}t]
                             \nonumber  \\
               &  &           -\frac{\gamma_-^2}{\Omega^2}\}e^{-\gamma_+t}.
\end {eqnarray}
The form of Eqn. ($10$) recalls Rabi' formula. According to the definition of
$E_-$, 'Resonance' means that the frequency of EM wave equals to the
energy difference including kinetic energy and internal energy of the
two unperturbed basis states divided by $\hbar$. While in most literatures,
resonance condition implies that the frequency of external field is tuned
the same as the transition frequency between two states of atom. So the
condition 'Resonance' used here includes the center-of-mass motion of the
atom. When the coupling between atom and EM field is strong
enough($\Omega>\gamma_-$), resonance condition $E_-/ \hbar+\Delta \omega=0$
guarantees the atom to oscillate between the state $\mid \varphi_1>$ and
$\mid \varphi_2>$ within a period $\Delta t<\tau_+=\frac{1}{\gamma_+}$.
This can be
clearly observed from Fig. $2$. At times $t>\tau_+$, the atom will decayed
into the lower energy states, since the probabilities of finding the atom
in $\mid\varphi_1>$ and $\mid \varphi_2>$
decrease rapidly with time for $t>\tau_+$.
Note that if the damping is excluded in our consideration,
i. e., $\gamma_-=\gamma_+=0$, then we recover the sinusoidal
oscillation as expressed in Eqn. $5$ in Ref. $13$.
In addition, according
to  Eqn. ($8$), EM wave provides  the atom
with a certain probability to transit from the lower
energy state $\mid \varphi_1>$ to the higher energy state $\mid \varphi_2>$
by absorbing a or several photons
of momenta $\hbar k$, before it decays to
lower energy states. If the EM wave is switched off, then
the atom will directly decay into other states than the
unperturbed state $\mid \varphi_2>$ and
it is impossible to find the atom in the higher energy states.
At $t=2n\pi/(\Omega^2-\gamma_-^2)^
{\frac12} (n=0,1,2,...)$, the probability of seeing the atom in state
$\mid \varphi_1>$
is zero, and the probability to find the atom still in
state $\mid \varphi_2>$ arrives maximum. For weak coupling,
the probabilities of finding the atom in its unperturbed states
is a sum of damped exponentials.
This result has a simple physical interpretation:
the lifetime  $ \tau_+=1/ \gamma_+$ is so short that
the system is damped into the lower energy states than
$\mid \varphi_2>$ before the coupling
has time to oscillate between states $\mid \varphi_2>$ and
$\mid \varphi_1>$.

\noindent
\begin{center}
{\bf V. Momentum Transfer between atom and EM Wave}
\end{center}

In our previous work, we have shown that a circular-polarized EM
wave exerts expelling or trapping force to a non-damped two-level
atom, due to the momentum transfer between the atom and EM wave.
The properties of the action depend on wether the atom motion
direction is the same as or opposite to the propagating direction
of EM wave, and the initial state of the non-damped atom. In this
section, we studied the momentum transfer between a damped atom
and a circularly polarized EM wave, and we only considered the
case that the atom is initially in its undamped state $\mid
\varphi_2>$ ($\gamma_2=0$). We also compared the results for
damped atom with that for non-damped atom. Using the expression of
probabilities to find the atom in its states $\mid \varphi_1>$ and
$\mid \varphi_2>$, we have the average values of the momenta  of
the damped atom
\begin{equation}
p_a=(p_0+\hbar k) \rho_1(p_0+\hbar k, t)+p_0 \rho_2(p_0,t)   .
\end {equation}
It is easy to show that  the expressions of averaged momenta
can be recovered into non-damped atom case in the absence of
damping emission. If the averaged momenta of
the atom is known, the momentum transfer between
atom and EM wave are calculated to be
\begin{equation}
\Delta p=p_a-p_0 .
\end {equation}
In the following, we discuss in detail the momentum  transfer
under the resonance condition. In the case of strong coupling, we have
\begin{equation}
\Delta p=(e^{-\gamma_1 t}-1)p_0 +\frac{\hbar k\Omega^2}{\Omega^2-\gamma_1^2}
\sin^2[(\Omega^2-\gamma_1^2)^{\frac12}t/2]e^{-\gamma_1t}+p_0
                            \frac{ \gamma_1(\Omega^2-\gamma^2_1)^{\frac12}}
                            {2\Omega^2}
                             \sin[(\Omega^2-\gamma^2_1)^{\frac12}t]e^{-
                             \gamma_1 t},
\end {equation}
for weak coupling, the momentum transfer is
\begin{equation}
\Delta p=(e^{-\gamma_1 t}-1)p_0 +\frac{\hbar k\Omega^2}{\gamma_1^2-\Omega^2}
\sinh^2[(\gamma_1^2-\Omega^2)^{\frac12}t/2]e^{-\gamma_1t}+p_0
                            \frac{ \gamma_1(\gamma^2_1-\Omega^2)^{\frac12}}
                            {2\Omega^2}
                             \sinh[(\gamma^2_1-\Omega^2)^{\frac12}t]e^{-
                             \gamma_1 t}.
\end {equation}
The average force exerted to the damped atom  by the EM wave can be also
obtained from Ehrenfest's theorem: $F=\frac{d}{dt} \Delta p$. From Eqn. (16)
, for $ \gamma_1 << \Omega $,  we can approximate
$e^{-\gamma_1 t}\simeq 1 $ and omit the first and third terms in Eqn. (16)
, then we recovered the results for non-damped atom: $\Delta p=
\hbar k \sin^2 (\Omega t/2)$ and $F=\Omega\hbar k \sin (\Omega t)/2$.
In this case, the action of the EM wave on the atom shows the expelling
property, because the direction of
the average momentum the atom requires from the
EM is the same as the propagating direction of the EM wave. This is easily
understood since the spontaneous emission has trivial effects on the
momentum transfer.  For $t<<\tau_1$, the results will be
in good agreement with that in Ref. 1.
Let us consider another limiting case $\tau_1>>\Omega$,
according to Eqn. (17),
 we can readily conclude that the momentum transfer in this limit
tends to zero rapidly. This phenomenon can
be understood easily since spontaneous emission induces
no net momentum transfer.
The most interesting is the
intermediate case where spontaneous emission process takes its effects.
As the transition probability, the momentum transfer
is expected to exhibit interesting oscillation in the strong coupling limit
and
rapid damping for weak coupling, though it is positive in the two limits.
When a damped atom is irradiated
by an EM wave, absorption-emission processes proceed at two distince rates.
Photons are absorbed  from and emitted into the applied wave at the induced
rate $\Omega$, and occasionally photons are spontaneous emitted
in random directions at the spontaneous rate $\gamma_1$. In the strong
coupling case that the induced rate exceeds the spontaneous rate, the
momentum transfer operating at the induced rate will be more
efficient and more  rapid than that at spontaneous rate.
Thus the momentum transfer will be similar to the case for
non-damped atom, with a resultant positive momentum transfer for a
short time showing
expelling property and a fluctuating ( the third term in Eqn. (16)
charactering the spontaneous process.   While for weak coupling,
the spontaneous rate exceeds the induced rate, then no net
momentum transfer is expected at times $t>\tau_1$, since the spontaneous
emission occurs in random directions and
thus introduces no net momentum transfer.

\noindent
\begin{center}
{\bf VI. Diagonization of \bf {H(t)} and 'Dressed States'}
\end{center}

    From Eqns. ($4$) and ($5$) or
following the relation ($20$) and ($21$) of Ref. $13$
between the amplitudes $\psi_1(p,t)$ and
$\psi_2(p,t)$, we can express
the wave function in the form
\begin{equation}
\Psi(P,t)=a_1\mid \varphi_-(t)>e^{-i[\frac{E_+}{2\hbar}+\beta^R(p)]}
e^{-[\frac
{\gamma_+}{2}-\beta^I(p)]t}
+a_2\mid \varphi_+(t)>e^{-i[\frac{E_+}{2\hbar}-\beta^R(p)]}
e^{-[\frac{\gamma_+
}{2}+\beta^I(p)]t}.
\end{equation}
where we defined a set of new time-dependent basis states $\mid
\varphi_-(t)>$
and $\mid \varphi_+(t)>$, which are the superposition of the unperturbed
basis states $\mid \varphi_1>$ and $\mid \varphi_2>$
\begin{eqnarray}
\mid \varphi_-(t)>&=&\cos(\vartheta_-)e^{-i\omega_Lt/2}
                     e^{i\theta_-} \mid \varphi_1>+\sin(\vartheta_-)
                      e^{i\omega_Lt/2}\mid \varphi_2>, \nonumber \\
\mid \varphi_+(t)>&=&\cos(\vartheta_+)e^{-i\omega_Lt/2}
                     e^{i\theta_+} \mid \varphi_1>+\sin(\vartheta_+)
                      e^{i\omega_Lt/2}\mid \varphi_2>,
\end {eqnarray}
with  $\vartheta_{\mp}=Arctg(\frac{\Omega}{2\mid \epsilon(p)
\mp\beta(p)\mid})$ and
$\theta_{\mp}=Arctg(\frac{\epsilon^I(p)\mp\beta^I(p)}
{\epsilon^R(p)\mp\beta^R(p)})$, where $\beta^R(p)$ and $\beta^I(p)$
are given by expression $(8)$.
In the $\{\mid\varphi_-(t)>,\mid
\varphi_+(t)>\}$ basis,
the total Hamiltonian $H(t)$ can be written as
\begin{equation}
H=\left[\begin{array}{cc} \frac{E_+}{2}+\hbar\beta^R(p)+
i\hbar(\gamma_+/2-\beta^I(p)) & 0 \\
0 &\frac{E_+}{2}-\hbar\beta^R(p)+i\hbar(\gamma_+/2+\beta^I(p))
\end{array} \right].
\end{equation}
Thus, in the $\{\mid \varphi_-(t)>, \mid \varphi_+(t)>\}$ space,
the total Hamiltonian of the system is diagonanized. The basis
states $\{\mid \varphi_-(t)>, \mid \varphi_+(t)>\}$ should be
regarded as the basis states belonging to the coupled system
between atom and EM wave, rather than the basis states of the
damped atom, since $\mid \varphi_-(t)>$ and $\mid \varphi_+(t)>$
include the contributions of the EM wave with circular
polarization, as can be seen from Eqn. ($15$). In this sense, we
call the states $\{\mid \varphi_-(t)>, \mid \varphi_+(t)>\}$
dressed states belonging to the coupled system, of which the
energies and the lifetimes are
\begin{eqnarray}
E_{\varphi_{\mp}}  &=&\frac12 [\frac{(p+\hbar k)^2}{2m}+\frac{p^2}{2m}\pm
                   \hbar\beta_R(p)],   \nonumber \\
\tau_{\varphi_{\mp}} &=& \frac{1}{\gamma_+\mp 2\beta^I(p)}.
\end{eqnarray}
For the above 'Dress states',
their lifetimes contain two parts, one of which is related to
the lifetimes of the atom in its unperturbed states, while another
depends on the interaction between atom and EM wave.
If we adjust the interaction between atom and EM field, we can
effectively elongate
the lifetime of the coupled system in the state $\mid \varphi_-(t)>$
and shorten the lifetime in $\mid \varphi_+(t)>$. In other words,
if an damped two-level atom is initially prepared in the dressed
state $\mid \varphi_-(t)>$, it would stay coherently in this state
for a longer time than $\tau_{\varphi_+}$,
since the spontaneous emission can be suppressed to a great extent.
This opens another way to prepare coherent atomic states.

Under the condition of resonance we have
\begin{eqnarray}
E_{\varphi_{\mp}}  &=&\frac12 [\frac{(p+\hbar k)^2}{2m}+\frac{p^2}{2m}\pm
                  \frac{\hbar(\Omega^2-\gamma_-^2)^{\frac12}}{2}],   \nonumber \\
\tau_{\varphi_{\mp}} &=& \frac{1}{\gamma_+},
\end{eqnarray}
for strong coupling ( $\Omega>\gamma_-$), and
\begin{eqnarray}
E_{\varphi_\mp}  &=&\frac12 [\frac{(p+\hbar k)^2}{2m}+\frac{p^2}{2m}],
                    \nonumber \\
\tau_{\varphi_{\mp}} &=& \frac{1}{\gamma_+\mp
                  (\gamma_-^2-\Omega^2)^{\frac12}}, \nonumber \\
\end{eqnarray}
for weak coupling  ( $\Omega<\gamma_-$).

From Eqns. $(22)$ and $(23)$, one can find that, if the EM wave is tuned
such that $E_-/ \hbar+\Delta \omega=0$ and the coupling is strong enough
($\Omega>\gamma_-$), then the dress states $\mid \varphi_-(t)>$ and
$\mid \varphi_+(t)>$ have different energies and the same lifetime;
in the weak coupling case ($\Omega<\gamma_-$), the energies
are the same and the lifetimes are different, this is to say,
for weak coupling, there exists degenerate dressed states for
the coupled system.

As a concrete example, we considered a moderately massive atom of
mass $m\simeq 10^{-23}$ $G$, transition frequency $\omega_a \simeq
10^{-16}$ $sec^{-1}$, original velocity $v_0\simeq 20$ $m
sec^{-1}$, lifetime of the excited state $\tau_1 \simeq
10^{-9}\sim 10^{-6}$ $ sec$, dipole momentum $D\simeq
 1.6 \times
10^{-29}$ $ C m$ and a circular polarized EM wave of angular
frequency $\omega_L \simeq \omega_a \simeq 10^{16}$  $sec^{-1}$.
Since $ p_0 \simeq 2\times 10^{-19}$ $ N sec<< mc\simeq
 3\times 10^{-12}$  $N sec^{-1}$,
the resonance condition is usually satisfied. To prolong the lifetime of
the dressed state $\mid \varphi_-(t)>$, the amplitude of
circular-polarized EM field should be small enough and must be
 less than the $10^{3} \sim 1$
 $V m^{-1}$.
Therefore if we want to prepare the dressed state mentioned above
at which the coupled system can keep for a long time, we may
irradiate the atom with a weak circular-polarized EM wave of the
same angular frequency as the atomic transition frequency. The
choice of the amplitude of EM wave depends on the lifetimes
(damping coefficients)  of the damped atom.

\noindent
\begin{center}
{\bf VI. Summary}
\end{center}

We have studied the dynamic effects of a circular-polarized EM wave on a
damped tow-level atom, with the consideration of the
center-of-mass motion of the atom. Schr$\ddot{o}$dinger equation of the
system and their strict solutions were given in Sec.II. It is shown that,
two unperturbed levels of the atom is split into
four levels of different energy and lifetime by the coupling between
the atom and EM wave. We found that the energies and lifetimes of the
four split-levels depend not only on the strength and frequency of the
EM wave, but also on the kinetic energy,damping rate and dipole moment
of the damped atom. The analytical relation between them was also presented.

  EM wave can introduces into the non-damped state as ground state part
  of the instability of the other states, hence no stable state exists for
  a damped atom interacting with  EM wave. When the frequency of the EM
  wave is tuned to satisfy the resonance condition, Rabi oscillation
  occurs in the transition probability for strong coupling, and rapid
  damping happens for weak coupling.

  Under resonance condition,
  momentum transfer between atom and EM wave shows
  interesting oscillating in the strong coupling case and rapid
  damping in the weak coupling case as transition probability,
  which depends on wether the induced emission-absorption rate
  or spontaneous rate exceeds the other. For $\gamma_1<<\Omega$,
  the momentum transfer for damped atom is very similar to that
  for non-damped atom, while for $\gamma_1>>\Omega$, the momentum
  transfer will tends to zero very rapidly.

   We also defined a new set of basis states-dressed states,
    which is the superposition of
  the unperturbed states and will evolve with time. In this new set of
    basis states, the total time-dependent Hamiltonian of the coupled
  system between the damped atom and EM wave is diagonanized and will
  be independent of time. Moreover, the energies and lifetimes of the
  new basis states can be effectively controlled. Thus the system can
  be coherently stay in one of the dressed states for a very long time, if
 the circular-polarized EM wave is tuned the same frequency
 and its amplitude is made appropriate according to the
 atomic damping coefficients.
 This opens another
 way to prepare coherent atomic states.

 It is interesting to compare the results with that without the
 consideration of damping in Ref. $13$, which studied
 the level split and momentum transfer between the atom and EM field,
 showing that the trapping or expelling
 force of EM wave on atom depends on the initial
 state of the atom.
 However, if spontaneous
 emission is taken into consideration, the energies of the split-levels
 will be related to the damping, and the probability
 of finding the atom in its all unperturbed states will not equal to
 $1$ as for non-damped atom. The momentum transfer is also related to
 the damping and the results for non-damped atom can be recovered as
 $\gamma_-<<\Omega$.
 Within a short period,
 the motion of the atom will be the same (sinusoidal oscillation)
 both for damped and for non-damped atom. In addition, the problem for
 damped atom is more practicable. Non-damped atom is only a
 particular case of damped atom. For damped atom, we can prepare
 a special state, in which the atom would stay coherently for
 a long time by the effective adjusting of the coupling.

 Due to the similarity between the quantum dots and atoms [$14$],
  some results in this paper
 can be directly applicable to the coupling of the quantum dots with
 an electromagnetic wave.

{\bf References}

\vspace{.1cm}
\noindent
$\ddagger$: zyzeng@mail.issp.ac.cn
\vspace{.1cm}
\noindent
\begin{enumerate}
\item R. J. Cook and A. F. Bernhardt, Phys. Rev. A {\bf18}, 2533 (1978).
\item J. Dalibard and C. Cohen-Tannoudji, J. Opt. Soc. Am. B {\bf6}, 2023
(1989).
\item M. H. Anderson et al., Science, {\bf269}, 198 (1995)
\item P. R. Berman, Cavity Quantum Electrodynamics (Academic press, INC,
Harcourt Brace \& Company, 1994).
\item E. M. purcell, phys. Rev. {\bf69}, 681  (1946).
\item D. Kleppner, Phys. Rev. Lett. {\bf47}, 237 (1981);
 S. Haroche and D. kleppner, Physics Today {\bf 42}, 24 (1989).
\item B. R. Mollow, Phys. Rev. {\bf 188}, 1969 (1969).
\item  C. Cohen-Tannoudji, J. Dupont-Roc, and G. Grynberg, Atom-Photon
Interactions (Willy, New York, 1992).
\item W. E. Lamb and R. C. Retherford, phys. Rev. {\bf79}, 549  (1950);
 ibid. {\bf81} 222 (1951).
\item H. A. Bethe and E. E. Salpetee, Quantum Mechanics of One- and Two-
Electron Atoms (Springer Verlag, 1957).
\item A. P. Kazantsev, G. J.Surdutovich, and V. P. Yakovlev, Mechanical
Action of Light on Atoms (World Scientific, Singapore 1990).
\item M. Wilkens, Phys. Rev. A {\bf 47}, 671 (1993).
\item Gao-Jian zeng, Shi-Lun zhou, Sheng-Mei Ao, and Zhao-Yang Zeng, Phys. Rev.
A {\bf55}, 2945 (1997).
\item L. Kouwenhoven and C. Marcus, Physics world, 35, June (1998).
\end{enumerate}

\vspace {5 cm}
\noindent

{\bf Figure Captions}

\vspace{.5cm}
\noindent

{\bf Fig.~1}~Ac Stark split of energy levels of a damped two-level
atom. \vspace{.5cm} \noindent

{\bf Fig.~2}~ Rabi oscillation (a) for strong coupling
 and damped exponential (b) for weak coupling  of transition
probability $\rho_1(P_0+\hbar k,t)$.

\end{document}